\documentclass{sf2a-conf2015}
\pdfoutput=1
\usepackage{graphicx}
\usepackage{hyperref}
\usepackage[]{natbib}  
\usepackage{epstopdf}

\def\BibTeX{{\rm B\kern-.05em{\sc i\kern-.025em b}\kern-.08em
    T\kern-.1667em\lower.7ex\hbox{E}\kern-.125emX}}
\bibpunct{(}{)}{;}{a}{}{,}  
\usepackage{amssymb, amsmath}
\usepackage{xspace}

\newcommand{\ramses}{{\textsc{Ramses}}\xspace}
\newcommand{\ramsesrt}{{\textsc{Ramses-RT}}\xspace}
\newcommand{\fesc}{{\ensuremath{f_{\text{esc}}}}\xspace}
\newcommand{\mvir}{{\ensuremath{M_{\text{vir}}}}\xspace}
\newcommand{\muv}{{\ensuremath{M_{\text{UV}}}}\xspace}
\newcommand{\Msun}{{\ensuremath{M_{\odot}}}\xspace}
\newcommand{\hi}{{\textsc{Hi}}}

\begin{document}

\TitreGlobal{SF2A 2015}


\title{Modeling small galaxies during the Epoch of Reionisation}

\runningtitle{Small galaxies during the reionisation}

\author{M. Trebitsch}\address{Universit\'e de Lyon, Lyon, F-69003, France;\\ Universit\'e Lyon~1, Observatoire de Lyon, 9 avenue Charles Andr\'e, Saint-Genis Laval, F-69230, France;\\ CNRS, UMR 5574, Centre de Recherche Astrophysique de Lyon;\\ \'Ecole Normale Sup\'erieure de Lyon, Lyon, F-69007, France}

\author{J. Blaizot$^1$}


\author{J. Rosdahl}\address{Leiden Observatory, Leiden University, P.O. Box 9513, 2300 RA Leiden, the Netherlands} 


\setcounter{page}{237}


\maketitle


\begin{abstract}
Small galaxies are thought to be the main contributors to the ionising budget of the Universe before reionisation was complete. There have been a number of numerical studies trying to quantify their ionising efficiency through the \emph{escape fraction} \fesc. While there is a clear trend that \fesc is higher for smaller haloes, there is a large scatter in the distribution of \fesc for a single halo mass. We propose that this is due to the intrinsic burstiness of star formation in low mass galaxies. We performed high resolution radiative hydrodynamics simulations with \ramsesrt to model the evolution of three galaxies and their ionising efficiency. We found that the variability of \fesc follows that of the star formation rate. We then discuss the consequences of this variability on the observability of such galaxies by JWST.
\end{abstract}



\section{Introduction}

One of the key science projects of the upcoming \emph{James Webb Space Telescope} (JWST) is to probe the end of the Dark Ages, around $z \sim 15$, when the first stars and galaxies formed. The apparition of these first sources of light marked the beginning of the Epoch of Reionisation (EoR), during which the matter in the Universe experienced a transition, from fully neutral to fully ionised. It is very appealing to link this phase transition to the formation of the first galaxies. Indeed, even if quasars produce much more ionising photons, the evolution of the quasar space density at high redshift seems to indicate that there are not enough of them to dominate the ionising background during the EoR  \citep{2015A&A...575L..16H}.

This favours a reionisation model in which galaxies are the main sources of ionising radiation. However, observational constraints from deep surveys such as the UDF12 campaign indicates that the galaxies detected by those surveys do not produce enough ionising radiation to reionise the Universe by $z \sim 6$ \citep{2013ApJ...768...71R}. This tension disappears if the luminosity function is extended to fainter galaxies, down to $M_{\text{UV}} \lesssim -13$. While these faint, low mass galaxies are thought to dominate the ionising budget of the Universe at high $z$, they have never been observed yet. Understanding their physical properties, formation histories and how ionising radiation can escape from them is therefore crucial to be able to prepare the next generation of deep surveys with the JWST.

In the past few years, there have been a number of numerical studies \citep[see e.g.][ for recent results]{2014ApJ...788..121K, 2014MNRAS.442.2560W, 2015MNRAS.451.2544P, 2015MNRAS.453..960M} trying to address the formation of those small galaxies, and to quantify the \emph{escape fraction} of ionising photons, which is the amount of radiation escaping from them. This typically requires very high resolution simulations, capable of resolving details in the insterstellar medium (ISM). Despite the variety of methods used for the modelling of star formation, supernova feedback, or radiative transfer, most of the recent simulations agree on the fact that the escape fraction \fesc decreases mass of the host dark matter halo $M_{\text{vir}}$. However, there is a large scatter in the \fesc vs. \mvir relation, that needs to be understood. We present here our contribution to the community effort to model those galaxies with cosmological simulations.
  
\section{Simulation methodology}

We performed cosmological simulations with the code \ramsesrt \citep{Rosdahl2013}, an extension of the cosmological code \ramses \citep{Teyssier2002} that solves the equations of radiative hydrodynamics (RHD) for astrophysical flows on an adaptive grid. \ramsesrt follows the coupled evolution of gas and radiation, allowing us to track the ionisation state of the gas in the simulation. We use the zoom technique to achieve the very high resolution needed to resolve the structure of the ISM. We focus on three haloes of masses $\mvir \sim 8 \times 10^7 \Msun$ for the smallest onne, $6 \times 10^8 \Msun$ for the intermediate one, and $2\times 10^9 \Msun$ for the largest one.

The haloes were selected in a dark-matter (DM) only simulation using $512^3$ particles in a $10 h^{-1}\ \text{Mpc}$ box. We then generated multigrid initial conditions using the \textsc{Music} code \citep{2011MNRAS.415.2101H}, using 3 additional levels of refinement, giving a DM particle mass of $\sim 2000 \Msun$. We ran the simulation for one billion year, down to $z \sim 5.6$, with a maximum of 21 levels of refinment, allowing for a cell size of $\Delta x \sim 7\ \text{pc}$. Radiation is modeled using three photon groups (ionising \hi, He \textsc{i}, He \textsc{ii}). We use the same supernova (SN) feedback recipe as in \citet{2014ApJ...788..121K}, where after 10 Myr, each star particle deposits in the ISM the amount momentum and metals corresponding to the phase of the SN explosion that can be resolved. We use a new recipe for star formation \citep{Devriendt}, where we account for the turbulence in the star-forming cloud.

\section{Bursty assembly of small galaxies}
\label{sec:bursty}

\begin{figure}[ht!]
 \centering
 \includegraphics[width=0.45\textwidth,clip]{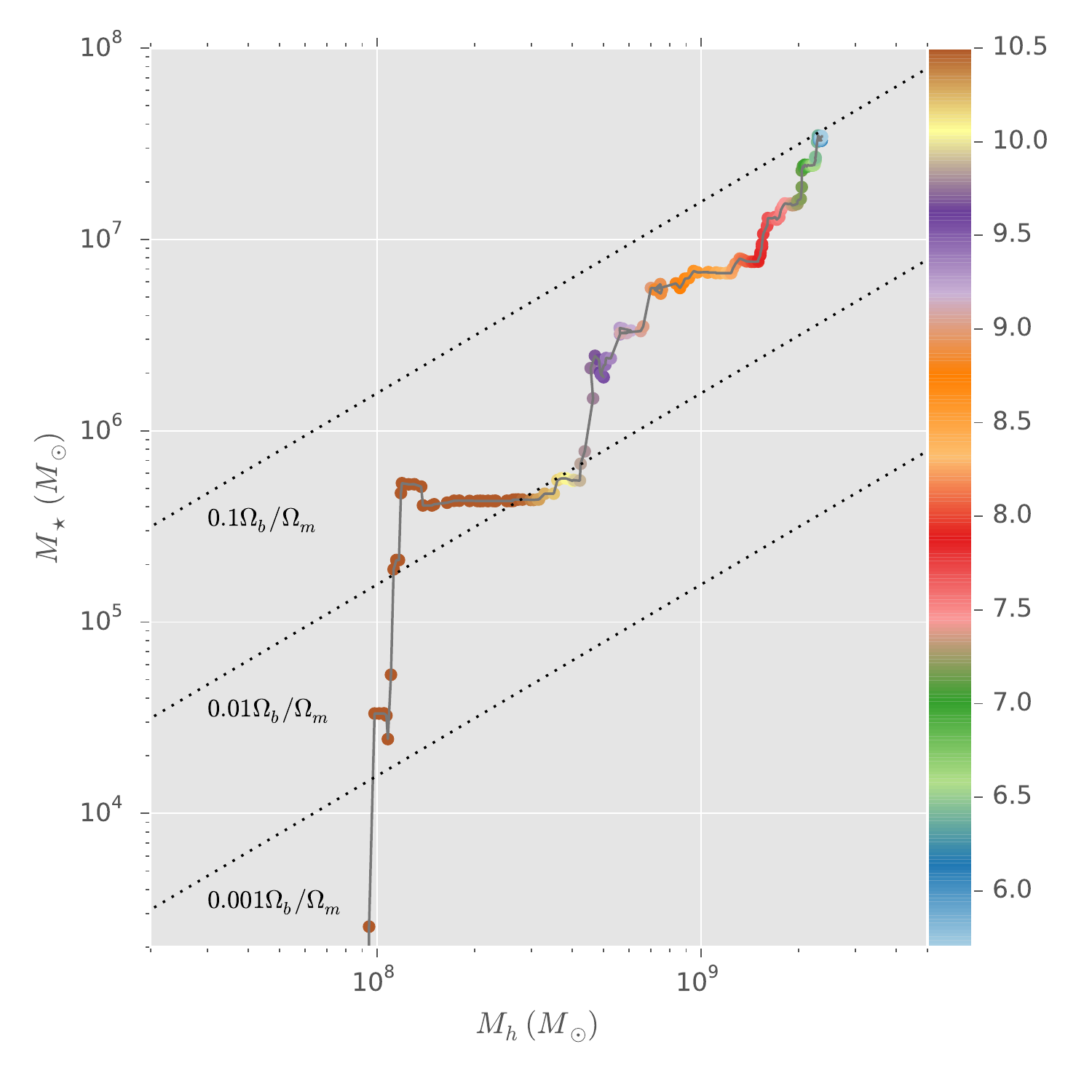}%
 \includegraphics[width=0.45\textwidth,clip]{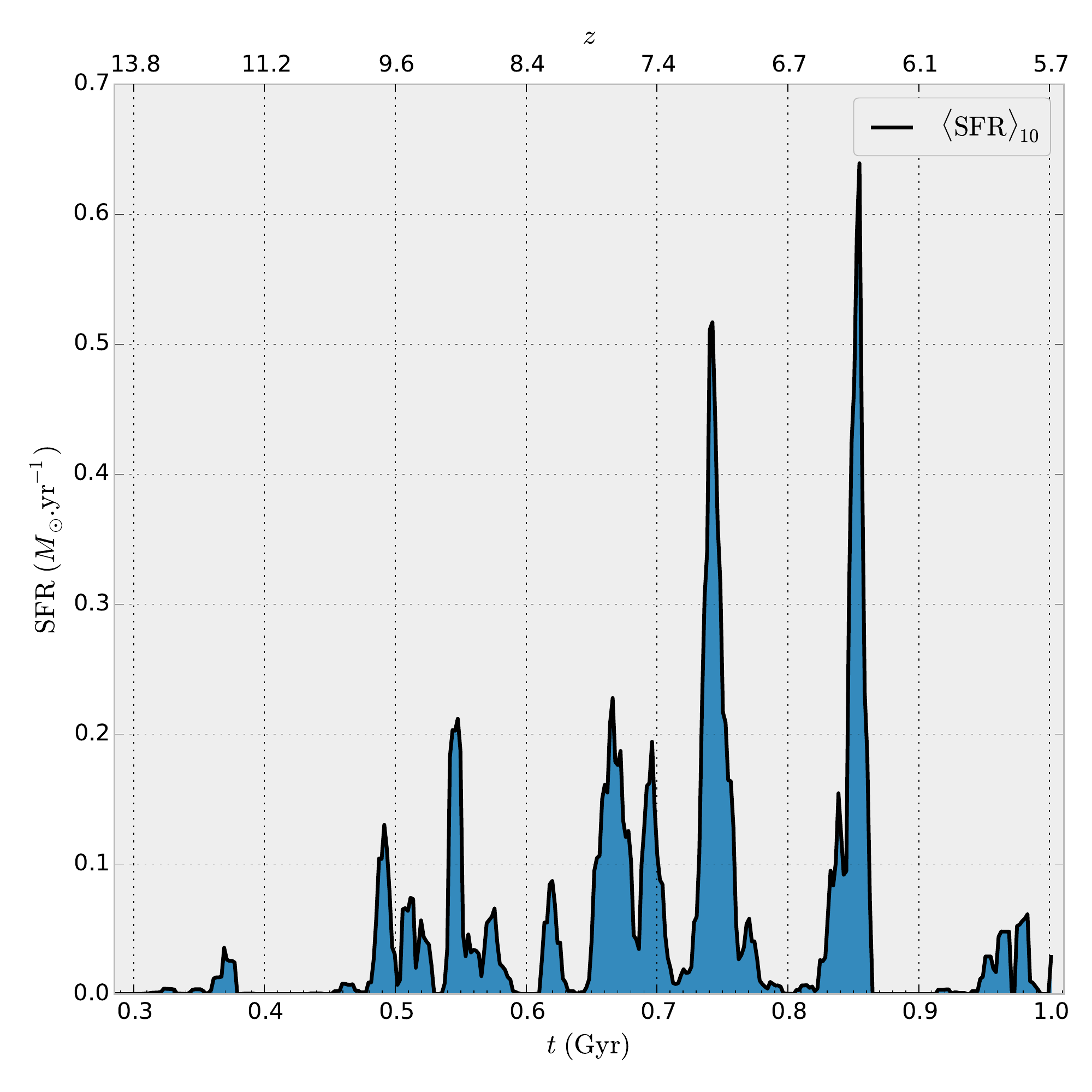}      
  \caption{{\bf Left:} Stellar mass to halo mass relationship for the most massive halo studied. The the dotted lines corresponds to a stellar fraction of 10\%, 1\% and 0.1\% of the baryons in the halo, and the colours note the redshift.  {\bf Right:} Star formation history of the galaxy. }
  \label{trebitsch:fig1}
\end{figure}

We present on Fig.~\ref{trebitsch:fig1} the assembly history of the most massive halo we targeted on our study. On the left panel, we show the evolution of the stellar mass to halo mass relationship with redshift, and on the right panel, we display the time evolution of the star formation rate (SFR) of the galaxy. In all the haloes, roughly between 1\% and 10\% of the baryons are converted into stars at all times, and a striking feature of the left panel is the presence of several plateaus indicating a growth of the galaxy with no associated star formation. The right panel provides a natural explanation for this: in this low mass regime, star formation happens by bursts. A few Myr after the beginning of a star formation episode, the most massive stars will end their life, and the resulting supernovae will heat and remove large amounts of gas from the ISM, quenching the star formation for a while. After some time, the gas will cool down in the halo, and return to the galaxy, fueling a new episode of star formation.

\begin{figure}[ht!]
 \centering
 \includegraphics[width=0.75\textwidth,clip]{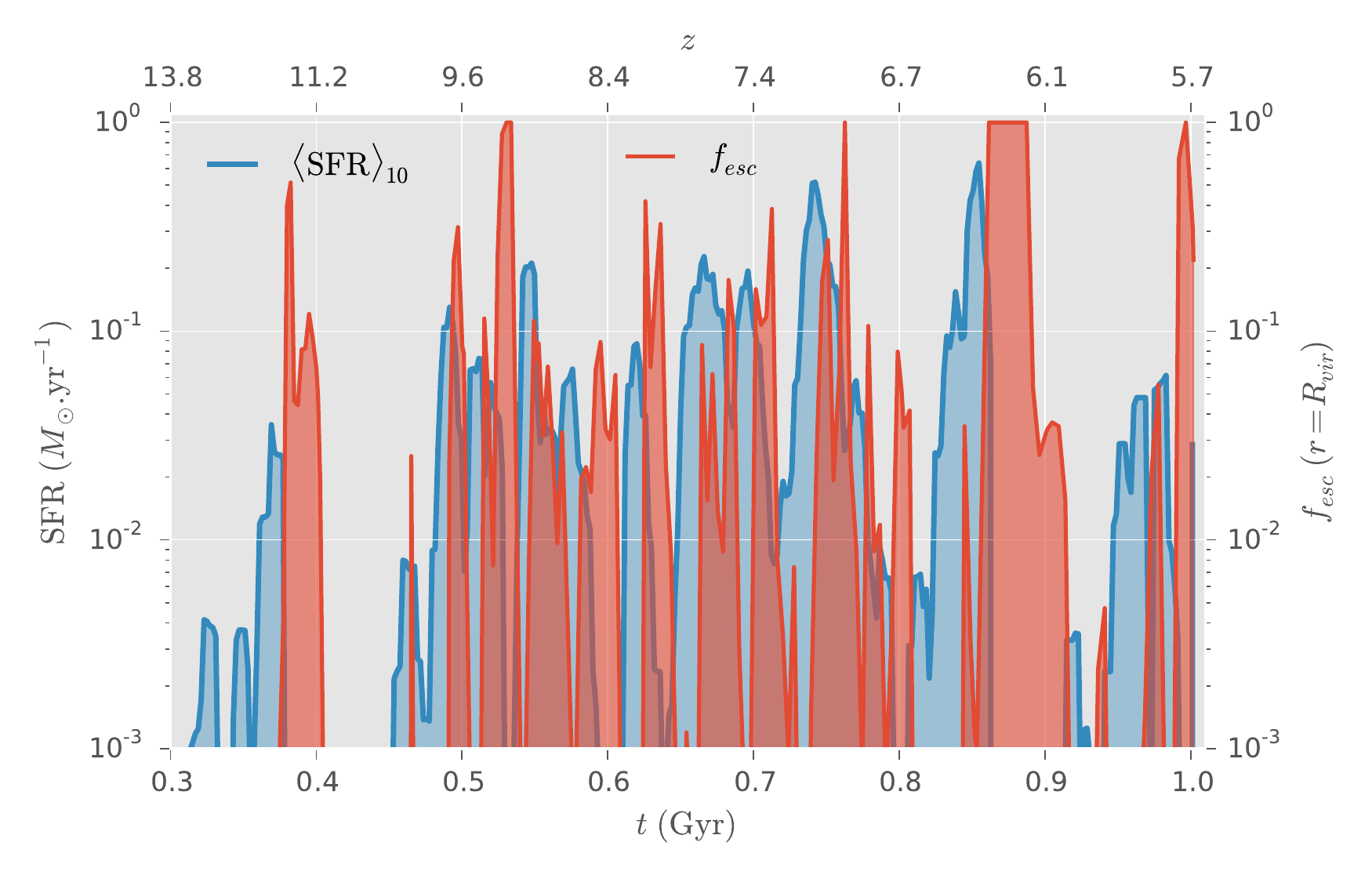}
  \caption{In blue, evolution of the SFR with time. The evolution of the escape fraction \fesc at the halo virial radius is show in red.}
  \label{trebitsch:fig2}
\end{figure}

This cycle of star formation episodes followed by outflow has very strong consequences for the escape of ionising radiation. Indeed, the supernovae will completely disrupt the star forming clouds in which most of the young stars live. This will vastly increase the ionising efficiency of the neighbouring young stars. We show on Fig.~\ref{trebitsch:fig2} the time evolution of both the SFR (in blue) and the escape fraction \fesc (in red). There is a clear correlation between the two curves, and \fesc starts to rise typically 10 Myr after the beginning of a star formation episode, which is the age at wich the star particles explode in supernovae in our simulations.

\section{Observational implications}
\label{sec:observations}

Using the models of \citet{Bruzual2003}, we computed the UV magnitude of the three galaxies we modelled to assess their observability with JWST. So far, our computations include neither dust nor IGM attenuation. We find that the most massive galaxy can reach absolute UV magnitudes as high as $\muv \lesssim -18$, but spend most of its time at $\muv \gtrsim -15$. We can expect that this kind of small, high-$z$ galaxy will be seen in deep JWST surveys, especially if they are in a starburst episode.

\begin{figure}[ht!]
 \centering
 \includegraphics[width=0.6\textwidth,clip]{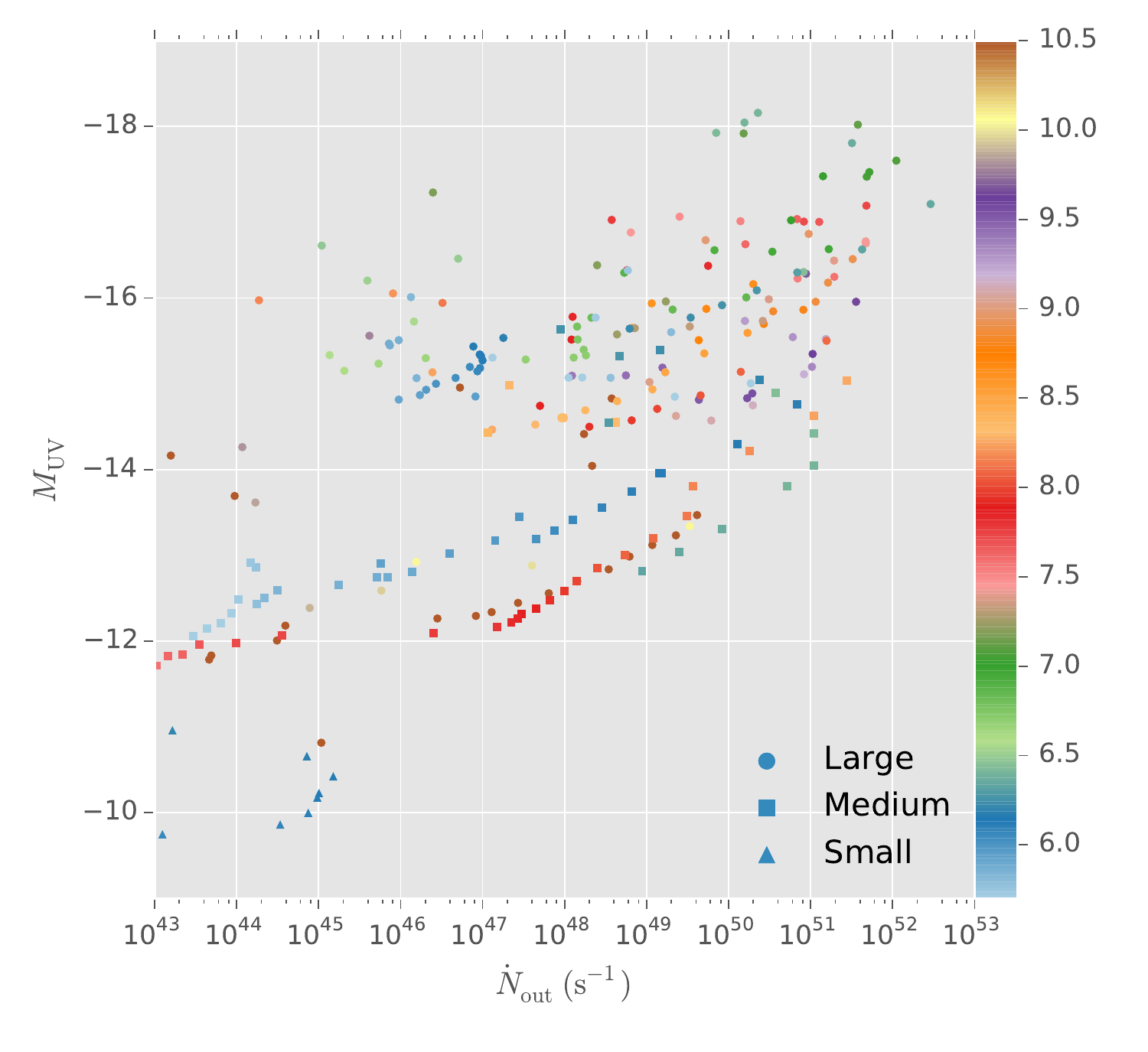}
  \caption{Ionising flux escaping the halo vs. UV magnitude. The different symbols represents the three different haloes, and the colours represent the redshift.}
  \label{trebitsch:fig3}
\end{figure}

We show on Fig.~\ref{trebitsch:fig3} the correlation between the UV magnitude and the ionising flux escaping the halo. While there is a clear trend that brighter galaxies are in general leaking more ionising photons, there is a large scatter. This means that selection only galaxies brighter than $\muv = -16$ will miss a large portion of galaxies that are actively contributing to the ionisation budget of the Universe. Conversely, a survey targeting the galaxies emitting more than $10^{50}$ photons per second would need to go deeper than $\muv = -14$ to be complete.

\section{Conclusions}

We have performed high resolution RHD simulations of three dwarf galaxies at $z \sim 6$ to investigate their ionising properties. We found that the cycle of star formation episodes followed by SN explosions will result in a highly varying escape of ionising radiation that can explain the scatter found in other studies. This highlights that it is critical to model star formation and feedback properly at small scale to accurately study the escape of ionising photons in high $z$ galaxies.

We then used simple SED modelling to assess the visibility of such galaxies. We found that the UV magnitude of these galaxies varies a lot with time (following the evolution of the SFR), between $\muv \sim -12$ and $\muv \sim -18$. Deep surveys with JWST will probably detect galaxies like these ones, but only when they are in a ``bright'' phase. We also found that there is only a loose correlation between UV magnitude and ionising emissivity.

\begin{acknowledgements}

MT and JB aknowledge support from the Lyon Institute of Origins under grant ANR-10-LABX-66, and from the ORAGE project from the Agence Nationale de la Recherche under grand ANR-14-CE33-0016-03. 
\end{acknowledgements}

\bibliographystyle{aa}  
\bibliography{trebitsch} 

\end{document}